\documentclass[epj]{svjour}
\usepackage{graphics}
\begin{document}
\title{A simple model for cavity-enhanced laser-driven ion acceleration
from thin foil targets}
\author{Piotr R\c{a}czka
\thanks{E-mail: piotr.raczka@ifpilm.pl}
\and Jan Badziak
}                     
%
%
\institute{Institute of Plasma Physics and Laser Microfusion, 23 Hery Str., 01-497 Warsaw, Poland}
\date{Received: date / Revised version: date}
%
\abstract{
A scenario for the enhanced laser-driven ion acceleration from a thin solid target at high laser intensity 
is considered, where the target is enclosed in a reflecting cavity. The laser pulse reflected from the target is redirected towards
the target by reflection at the inner cavity wall. This scenario is discussed in the context of sub-wavelength foil acceleration in the radiation
pressure regime, when plasma dynamics is known to be reasonably well
described by the light-sail model. A semi-analytic formula is presented that describes the effect of cavity reflections in one-dimensional geometry.  The effect of cavity reflections
on sub-wavelength foil acceleration is then evaluated for two concrete
examples of ultra-intense laser pulses of picosecond and femtosecond duration.
\PACS{
      {52.38.Kd}{Laser-plasma acceleration of electrons and ions}   \and
      {41.75.Jv}{Laser-driven acceleration}
     } 
} 

\maketitle
%
\section{Introduction}
\label{intro}
In the last decade there has been growing interest in laser-driven
ion acceleration, where energetic ions are generated by irradiating
a thin foil (with thickness in the micrometer range) with an ultra-intense
(above $10^{18}$~W/cm\textsuperscript{2}) laser beam. The laser-generated
proton pulses had been used for example for ultra-fast radiography
and radioisotope production, and together with laser-generated beams
of heavier ions they are of high interest from the point of hadron
tumor therapy and fast ignition of inertial confinement fusion targets
\cite{06-RMP-78p309-Mourou,06-FST-49p412-Borghesi,06-NP-2p48-Fuchs,07-NP-3p58-Robson}.
The use of an ultra-intense laser as an ion driver offers certain
advantages over conventional linear or circular accelerators, allowing
for construction of very compact devices that produce particle pulses
of high intensity and very short duration. Unfortunately, ion energies
achieved so far in this approach are not very high (below 100~MeV
per nucleon), and the achieved values of the laser-to-ion energy conversion
efficiency are generally below 10\% \cite{11-PRL-107p115002-Jung,11-NF-51p083011-Hegelich}.

\begin{figure}
\centering
\resizebox{0.75\columnwidth}{!}{\includegraphics{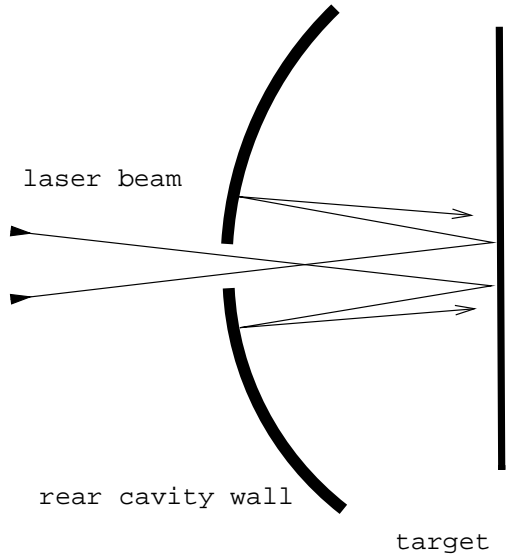}}
\caption{A schematic drawing of a target enclosed in a cavity that redirects the pulse reflected from the foil back into the foil.}
\label{fig:cavity target}
\end{figure}

It was recently observed that basic parameters of the laser-accelerated
ion pulses - the maximum ion energy, the average ion energy, and the
laser-to-ion energy conversion efficiency - could be considerably
improved if the target foil is enclosed in a cavity that redirects
the laser pulse reflected from the foil back into the foil, thus in
a sense ``recycling'' the laser energy \cite{10-APL-96p251502-Badziak,11-EPS-P5.038-Badziak,12-PP-Badziak}.
This mechanism, dubbed the Laser
Induced Cavity Pressure Acceleration (LICPA), was found
to be very effective at sub-relativistic laser intensities ($\sim10^{15}$~W/cm\textsuperscript{2}),
where enhancement in ion acceleration is predominantly due to the
buildup of thermal plasma pressure inside the cavity \cite{10-APL-96p251502-Badziak,12-PP-Badziak,09-APL-95p231501-Borodziuk}.
Computer simulations performed at higher laser intensities indicate that the LICPA mechanism
would be effective also in the case when the radiation pressure becomes important  \cite{11-EPS-P5.038-Badziak,12-PP-Badziak,12-APL-101p084102-Badziak}.
A possible target arrangement realizing the LICPA principle in the
radiation pressure regime is shown in Fig.~\ref{fig:cavity target}. The validity of the cavity-enhancement idea is confirmed by a recent experiment \cite{12-APL-101p024101-Scott}, where a target with a half-cavity of spherical shape was used to redirect and refocus the reflected pulse at the intensity of $10^{19}$~W/cm$^2$.
In this note we consider a realization of the LICPA idea in the radiation
pressure regime, but in a narrower context, namely when the plasma dynamics
is reasonably well described by the light-sail (LS) model \cite{66-N-211p22-Marx,93-AJP-61p205-Simmons,04-PRL-92p175003-Esirkepov}.
Such a situation occurs for example when a high intensity and high
contrast laser pulse is incident on a foil of sub-wavelength thickness.
This acceleration regime attracted recently considerable attention
\cite{07-PRL-99p065002-Pegoraro,08-NJP-10p013021-Robinson,08-PRST-AB-11p031301-Klimo,08-PPCF-50p124033-Liesykina,09-PPCF-51p024014-Tripathi,09-CRP-10p216-Bulanov,09-PRL-102p145002-Qiao,09-PRL-103p085003-Macchi,10-NJP-12p045013-Macchi}, because it may lead to creation of bunches of highly energetic ions with low energy dispersion.
 We point out that in the LS regime and in a one-dimensional approximation one may  obtain a semi-analytic description of the cavity-enhanced ion acceleration by adapting observations on LS dynamics made in \cite{93-AJP-61p205-Simmons}.  We employ  this semi-analytic description to discuss some characteristic features of the cavity-enhanced ion acceleration for  realistic  ultra-thin foil targets and laser parameters
close to the expected capabilities of the big new laser facilities
that should become available in the near future \cite{08-PPCF-50p124045-Blanchot,11-ELI-Whitebook}.

\section{The light-sail model in one dimension}
\label{sec:1}
Let us consider a thin planar foil of areal mass density $\sigma$,
irradiated at  normal incidence by an intense laser beam propagating
along the $x$ axis, with the temporal profile of intensity $I(t)$ in the laboratory reference frame. Let us also assume that the foil is ideally reflective in its instantaneous rest frame.
The equation of motion
for such a foil driven by the laser beam had been first derived
in the context of laser-driven interstellar space travel \cite{66-N-211p22-Marx,93-AJP-61p205-Simmons}. The solution of this equation for the speed of the foil which was at rest at $t=0$ has the following form:
%
\begin{equation}
\beta(w)=\frac{(1+e(w))^{2}-1}{(1+e(w))^{2}+1}\,,\label{eq:beta-solution}
\end{equation}
where $\beta=\frac{1}{c}\frac{dx}{dt}$, $w=ct-x$ is the retarded relativistic time variable, and  $e(w)$ is the total laser energy per unit area incident on
the foil up to the retarded time $w$, normalized to the half of the
relativistic rest energy per unit area of the foil:
\begin{equation}
e(w)=\frac{2}{\sigma c^{3}}\int_{0}^{w}I(w'/c)dw'\,.\label{eq:en-dep}
\end{equation}
%
%
This very simple 1D model was found to be surprisingly useful in interpreting
results of 3D and 2D kinetic simulations for laser ion acceleration.
The  light-sail behaviour was famously identified in a 3D simulation
at the laser intensity of $1.4\times10^{23}$~W/cm\textsuperscript{2}
\cite{04-PRL-92p175003-Esirkepov}. It was found that the initially
planar foil is deformed into a ``cocoon'', in which the laser pulse
is confined, and a stable plasma clump is formed from the portion
of the foil, that is being then pushed
forward by the radiation pressure.
 In simulation reported in \cite{04-PRL-92p175003-Esirkepov} a linear polarization was assumed for the laser beam, but it it
was later observed that circularly polarized beams create more favorable
conditions for this mode of plasma acceleration \cite{05-PRL-94p165003-Macchi},
and a parametric study in 1D had shown that the LS-type of behavior becomes
visible for ultra-thin targets already at $I\times(\lambda/1\mu\textrm{m})^{2}=2.8\times10^{20}$~W/cm\textsuperscript{2} \cite{09-PRL-103p085003-Macchi}.
The case of linear polarization is more complex, because generally
at less extreme intensities a broad energy spectrum is obtained.
In the following we shall assume that the laser beam is circularly polarized.
It is interesting that the evolution of targets which
are relatively thick compared to the laser wavelength is also found
to be consistent with the LS dynamics \cite{11-NJP-13p123003-Grech,11-APL-99p071502-Badziak}, although such targets are not accelerated as rigid bodies and undergo internal evolution.


A generalization of the relativistic LS foil dynamics beyond the one-dimensional approximation  was
outlined in \cite{07-PRL-99p065002-Pegoraro,09-CRP-10p216-Bulanov}
and applied to the study of instabilities appearing in the acceleration of  ultra-thin foils.

\section{Cavity-enhanced light-sail acceleration}
Let us now consider foil acceleration with cavity enhancement, as illustrated in Fig.~\ref{fig:cavity target}. When a circularly polarized beam is incident on a target foil that is sufficiently thick to be opaque to the laser light, a substantial part of the beam is reflected from the foil and lost from the system. When the target foil is enclosed in a cavity with a massive rear wall, this reflected light  is reflected once again at the rear cavity wall and redirected into the foil. One could imagine that for a properly designed cavity it may be possible for the laser light to circulate back and forth inside the cavity several times. This could result in a significant enhancement of the ion acceleration from the foil in comparison with the standard arrangement.

The interaction of the laser beam with the foil and the cavity is of course a complicated multi-scale phenomenon, which would be studied through detailed numerical simulations elsewhere. In this note we want to discuss a very simple model of this interaction, which nevertheless may be argued to be sufficiently realistic to properly illustrate some essential features of the cavity-enhanced acceleration.  The big advantage of of this model is that it may be described by a semi-analytic formula, which makes it easy to perform all kinds of parametric studies.

The basic assumptions of the model are as follows: (1) the dynamics is one-dimensional (which may be realized through an optimized choice of the shape of the cavity, but is of course a crude approximation due to the evolution of the foil and the  divergence of the laser beam); (2) during the acceleration the foil remains ideally reflective in its instantaneous rest frame (although during the evolution of the target normal incidence could not be preserved on all segments of the foil, so that even for a circularly polarized beam some laser energy would be lost due to the heating of the target); (3) the fraction $R_{c}$ of the reflected laser energy being redirected back into the target does not change as a result of evolution (in reality $R_{c}$ may be affected by the deteriorating quality of the reflected light, resulting in enhanced heating of the rear cavity wall, as well as interactions with side walls of the cavity and the low-density plasma expanding into the cavity; however, our experience with cavity-enhanced dynamics shows that only first few reflections are of practical interest).

The parameter $R_{c}$ is of crucial importance in the cavity-enhanced acceleration. It is difficult to give an \emph{a priori} estimate for $R_{c}$, because it is primarily affected by the amount of the laser light being lost through the entrance to the cavity, which in turn is strongly dependent on  the 3D geometry of the concrete cavity and the transverse profile of the laser beam.   In further calculations we assume that in proper conditions for circular polarization $R_{c}$ could have values in the range $0.5\div 0.7$.

The problem of generalization of the 1D LS dynamics to take into account the contribution from a recirculating pulse was considered already
in \cite{93-AJP-61p205-Simmons}
in the context of the interstellar space travel driven by an Earth-based
laser. In the following we adapt the reasoning presented in  \cite{93-AJP-61p205-Simmons} to the case of cavity-enhanced laser-driven acceleration in the microscale: we write down an explicit formula that is valid for arbitrary temporal profile of the driving beam and allows for determination of the foil position at arbitrary time.
%

Let us assume that the laser pulse enters the cavity at $x=0$ and
is
reflected by the foil located at $x(w)$.
The pulse returns to the inner cavity
wall located at $x=0$ and is then redirected towards the foil.
Let us denote by $w_{j}$, $j=0,1,2....$,
the values of the retarded time for which the leading edge of the
laser pulse is redirected towards the foil from the position of the inner
cavity wall for the $j$-th time. Assuming that the laser pulse starts
at $w_{0}=0$ and the foil is initially at $x(0)=L_{c}$, where $L_{c}$ denotes the depth of the cavity, we obviously have $w_{1}=2L_{c}$. Next, let $x^{(j)}(w)$
denote (for $j\geq1$) the position of the foil for $w$ in the interval
$[w_{j-1},w_{j}]$. These functions satisfy an obvious continuity
relation $x^{(j+1)}(w_{j})=x^{(j)}(w_{j})$. Given the values of $w_{j}$ and
$x^{(j)}(w_{j})$, the value of $w_{j+1}$ is fixed by the formula
\begin{equation}
w_{j+1}=w_{j}+2x^{(j)}(w_{j})\,.
\end{equation}
Finally, let $e^{(j)}(w)$ denote (for $j\geq1$) the total laser
energy per unit area of the foil - expressed in the units of a half
of the relativistic rest energy per unit area of the foil - that was
incident on the foil for $w$ in the interval $[w_{0},w_{j}]$, i.e.
during the whole acceleration process up to $w_{j}$. The value of
$e^{(1)}(w)$ is simply given by Eq.~\ref{eq:en-dep}. The
values of $e^{(j)}(w)$ for $j\geq2$ may be determined iteratively.
The expression for $e^{(j+1)}(w)$ consists of three terms: the first
representing the total (normalized) laser energy per unit area incident on the
foil up to $w=w_{j}$, the second representing the energy deposited by the primary laser pulse
in the interval $[w_{j},w]$, and the third representing the 
energy per unit area that had been reflected from the foil in the
interval $[w_{j-1},w_{j}]$ and was then redirected towards the foil. The
laser energy per unit area reflected from the foil in an interval
$[w_{a},w_{b}]$ is given by the formula
\begin{equation}
\frac{\sigma c^{2}}{2}\frac{e(w_{b})-e(w_{a})}{[1+e(w_{b})][1+e(w_{a})]}\,,
\end{equation}
but only a fraction $R_{c}$ of this is redirected to the foil by
the reflection at the inner cavity wall.  Taking into account all contributions we
obtain
\begin{equation}
e^{(j+1)}(w)=e^{(j)}(w_{j})+e^{(j+1)}_{dir}(w)+e^{(j+1)}_{recirc}(w),
\label{eq:cavity-en-dep-tot}
\end{equation}
where
\begin{equation}
e^{(j+1)}_{dir}(w)=\frac{2}{\sigma c^{3}}\int_{w_{j}}^{w}I(w'/c)dw'
\label{eq:cavity-en-dep-dir}
\end{equation}
represents the effect of deposition of the laser energy by the primary laser pulse and
\begin{equation}
e^{(j+1)}_{recirc}(w)=R_{c}\frac{e^{(j)}(w_{ref}^{(j)}(w))-e^{(j)}(w_{j-1})}{[1+e^{(j)}(w_{ref}^{(j)}(w))][1+e^{(j)}(w_{j-1})]}\,,
\label{eq:cavity-en-dep-recirc}
\end{equation}
represents the effect of recirculated laser energy. The function $w_{ref}^{(j)}(w)$ appearing above is a solution of an implicit
equation
\begin{equation}
w=w_{ref}^{(j)}+2x^{(j)}(w_{ref}^{(j)})\,,
\end{equation}
i.e. it represents the value of the retarded time variable from the
interval $[w_{j-1},w_{j}]$ for which a redirected ray that is incident
on the foil at $w\in[w_{j},w_{j+1}]$ was for the last time reflected
from the foil. Obviously, $w_{ref}^{(j)}(w_{j})=w_{j-1}$. The relativistic
speed $\beta^{(j+1)}(w)$ in the interval $w\in[w_{j},w_{j+1}]$ is
obtained by substituting $e^{(j+1)}(w)$ into  Eq.~\ref{eq:beta-solution}.
The position of the foil is then obtained from the formula
\begin{equation}
x^{(j+1)}(w)=x^{(j)}(w_{j})+\int_{w_{j}}^{w}\frac{\beta^{(j+1)}(w')}{1-\beta^{(j+1)}(w')}dw'\,.
\end{equation}
 The relativistic kinetic energy of the accelerated ions for $w\in[w_{0},w_{j}]$
is given by the formula
\begin{equation}
E_{i}(w)=m_{i}c^{2}\frac{(e^{(j)}(w))^{2}}{2[1+e^{(j)}(w)]}\,.\label{eq:ion-kin-en}
\end{equation}
If we use Eq.~\ref{eq:cavity-en-dep-tot}, Eq.~\ref{eq:cavity-en-dep-dir}, and Eq.~\ref{eq:cavity-en-dep-recirc} to calculate $e^{(j)}(w)$ at $w=w_{j}$, assuming  a  rectangular temporal profile of the pulse and  $R_{c}=1$, we obtain expression given in Eq.~(13a) in \cite{93-AJP-61p205-Simmons}.

Despite its simplicity and highly approximate character
 this model is useful in providing some concrete numbers
for cavity-enhanced ion acceleration at future high-intensity laser facilities.

\section{Concrete examples}
Let us consider the predictions of the cavity-enhanced LS model in
two concrete scenarios characterized by the same total laser fluence,
but with distinctly different pulse lengths.
The first scenario involves a pulse with the maximum intensity of $10^{21}$~W/cm$^{2}$ and the full width at half maximum equal to $t_{HW}=1.3$~ps; such parameters should be in principle available
for example at the future PETAL facility \cite{08-PPCF-50p124045-Blanchot}.
The second scenario involves
a laser pulse with the maximum intensity of $10^{22}$~W/cm$^{2}$ and duration of 130~fs; such
parameters should be in principle attainable for example with a slightly  defocused
kilojoule beam at the future ELI Beamlines facility \cite{11-ELI-Whitebook}.
The temporal profiles of these pulses are taken to be super-Gaussian,
and they are assumed to start at the intensity smaller by a factor
of 100 relative to the maximum intensity: $I(t)=I_{0}\exp[-((t-t_{p}/2)/\tau)^{6}]$
for $t\in[0,t_{p}]$, where $t_{p}=t_{HW}(\log100/\log2)^{1/6}$ and
$\tau=\frac{1}{2}t_{HW}/(\log2)^{1/6}$. The laser energy fluence
on target in both scenarios is equal to 12.81~J/($\mu$m)$^{2}$.

Our target is assumed to be an ultra-thin carbon foil, with the density
of $2.2$~g/cm$^{3}$.  In 1D the condition for nearly total reflection
takes the form \cite{09-PRL-103p085003-Macchi,10-NJP-12p045013-Macchi}
$\pi(l/\lambda)\times(n_{e}/n_{c})>a_{0}$, where $l$ is the foil thickness, $\lambda$ is the
laser wavelength (which is either 800~nm for ELI or 1054~nm for
an Nd:glass picosecond laser), $n_{e}$ is the electron number density in the ionized target,
$n_{c}=(4\pi^{2}c^{2}m_{e}\varepsilon_{0}/e^{2})\times(1/\lambda^{2})$
is the critical electron density for a given wavelength ($m_{e}$
is the electron mass), and $a_{0}$ is the dimensionless laser field
amplitude, related to the intensity $I$ of a circularly polarized beam by the relation $I= m_{e}c^{3}n_{c}a_{0}^{2}$.
The non-transparency condition may be rewritten in
the form
\begin{equation}
l>\frac{2}{en_{e}}\sqrt{\frac{\varepsilon_{0}I}{c}}.\label{eq:non-transp}
\end{equation}
For a fully ionized carbon we have $n_{e}=6.62\times10^{23}$~cm\textsuperscript{-3},
so for a circularly polarized beam with $I=10^{22}$~W/cm$^{2}$ we obtain a condition $l>32$~nm.
In the following we present concrete predictions for $l=200$~nm,  which in our view is a realistic value, leaving  some margin for possible prepulse issues which might be important
in the case of kilojoule beams. We then discuss how the achieved ion energies would scale with the thickness of the target.
As for the cavity parameters, in both cases we choose as a reference value $L_{c}=80$~$\mu$m,
which seems to be a technologically achievable value, and we discuss the $L_{c}$ dependence separately.

\begin{figure}
\centering
\resizebox{0.75\columnwidth}{!}{%
\includegraphics{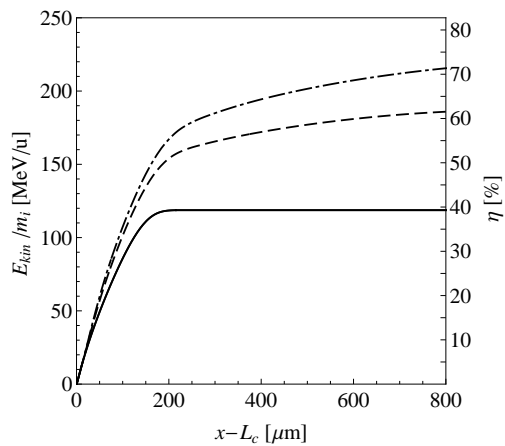}
}
\caption{The ion energy per a.m.u., as a function of the acceleration
distance, for the case of 1.3~ps pulse with maximum intensity of 10\textsuperscript{21}~W/cm\textsuperscript{2}peak. The solid line indicates acceleration without cavity reflection,
while the dash-dotted (dashed) line indicates an acceleration with
the fraction $R_{c}=0.7$ ($R_{c}=0.5$) of the laser energy being
reflected at the inner cavity wall.}
\label{fig:petal pulse}
\end{figure}

In Fig.~\ref{fig:petal pulse} we show the evolution of the ion energy
per atomic mass unit of ion mass (i.e. $E_{ion}^{kin}/m_{i}$, where $m_{i}$
is the ion mass) as a function of the acceleration distance
for the 1.3~ps pulse with 10\textsuperscript{21}~W/cm\textsuperscript{2} peak
intensity. The evolution in the absence of the cavity reflections
is indicated by the solid line: the foil is rapidly accelerated over
a distance of 215~$\mu$m to the energy of 119~MeV/u, and then ``sails
away'' at a constant speed.

When reflections at the inner cavity wall are taken into account
 the acceleration of the foil is continuously enhanced
by the laser pulse redirected multiple times towards the target by
reflection at the inner cavity wall (as indicated by the dash-dotted
line). For  $R_{c}=0.7$  the leading edge of the laser pulse hits the accelerating foil for the fourth time when the foil has moved approximately $410$~$\mu$m from the starting position. The recirculation of laser light results in nearly doubling of the resulting ion energy relative to the
case without the cavity reflection. For $R_{c}=0.5$ (dashed line)
the energies to which the ions are accelerated are slightly smaller,
but the general pattern is the same.

The result for the ion energy per a.m.u. is easily converted into
the laser-to-ion conversion efficiency, which is shown on the right
side of the figure. The total ion energy per unit surface area of
the target is given by $\sigma \times E_{ion}^{kin}/m_{i}$, so every 10~MeV of ion energy per a.m.u.
corresponds to the ion beam energy fluence of $42.45$~MJ/cm\textsuperscript{2},
which implies a laser-to-ion energy conversion efficiency of 3.31\%.
Note that the acceleration efficiency is quite high - around 40\%
- even without the cavity reflections, which is characteristic of
the radiation pressure acceleration in 1D geometry \cite{04-PRL-92p175003-Esirkepov,09-PRL-103p085003-Macchi}.

\begin{figure}
\centering
\resizebox{0.75\columnwidth}{!}{%
\includegraphics{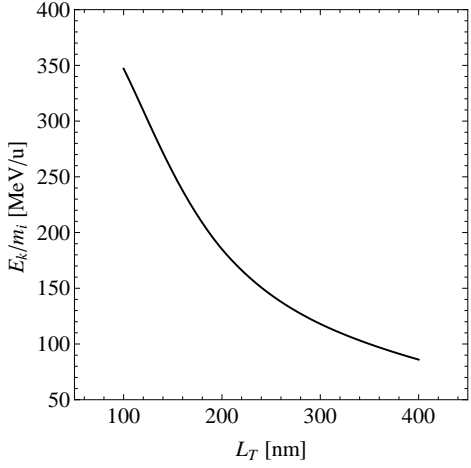}
}
\caption{The ion energy per a.m.u. at the acceleration distance of 300~$\mu$m, as a function of the thickness of the target foil,  for the
1.3~ps pulse with intensity of 10\textsuperscript{21}~W/cm\textsuperscript{2} peak, with the fraction $R_{c}=0.7$ of the laser energy being
reflected at the inner cavity wall. }
\label{fig:petal-pulse-eltarg}
\end{figure}

The results for ion energies reported above correspond to the target foil thickness of $l=200$~nm, but due to the simplicity of the LS dynamics it is straightforward to extend them to other values of this parameter. In Fig.~\ref{fig:petal-pulse-eltarg} we show the ion energy per a.m.u. at the acceleration distance of 300~$\mu$m, as a function of thickness of the target foil, with the fraction $R_{c}=0.7$ of the laser energy being
reflected at the inner cavity wall. For thinner targets higher ion energies are achieved, although the relation is far from trivial, since lighter targets are moving faster and may experience smaller number of reflections. However, in practice very thin targets may be used only when the level of contrast in the pulse is high, so that the foil would not damaged by the prepulse. It is interesting that energies of the order of 200~MeV/u - which is considered to be an important threshold for example for the hadron therapy~\cite{06-RMP-78p309-Mourou} - seem to become accessible in the arrangement considered here.

\begin{figure}
\centering
\resizebox{0.75\columnwidth}{!}{%
\includegraphics{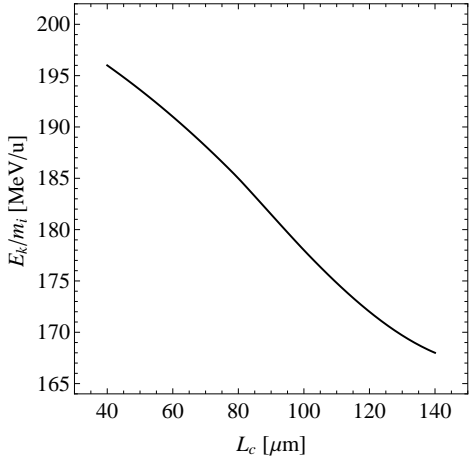}
}
\caption{The ion energy per a.m.u. at the acceleration distance of 300~$\mu$m, as a function of the depth $L_{c}$ of the cavity,  for the
1.3~ps pulse with intensity of 10\textsuperscript{21}~W/cm\textsuperscript{2} peak, with $R_{c}=0.7$ of the laser energy being
reflected at the inner cavity wall.  }
\label{fig:petal-pulse-elcav}
\end{figure}

In calculations shown in Fig.~\ref{fig:petal pulse} we assumed $L_{c}=80$~$\mu$m. In Fig.~\ref{fig:petal-pulse-elcav} we show the ion energy per a.m.u. at the acceleration distance of 300~$\mu$m, as a function of the depth $L_{c}$ of the cavity,   with $R_{c}=0.7$. Somewhat surprisingly,  this dependence turns out to be  rather weak.

\begin{figure}
\centering
\resizebox{0.75\columnwidth}{!}{%
\includegraphics{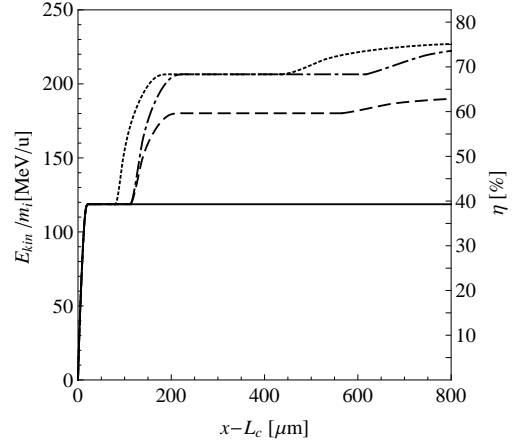}
}
\caption{The ion energy per a.m.u., as a function of the acceleration
distance, for the 130~fs pulse with peak intensity of 10\textsuperscript{22}~W/cm\textsuperscript{2}. The solid line indicates acceleration without cavity reflections,
while the dash-dotted (dashed) line indicates the acceleration with
the fraction $R_{c}=0.7$ ($R_{c}=0.5$) of the laser energy being
reflected at the inner cavity wall. The dotted line corresponds to the case of a shorter cavity with $L_{c}=60$~$\mu$m.}
\label{fig:eli-pulse}
\end{figure}

The evolution of ion energy per a.m.u. for the case of 130~fs pulse
with 10\textsuperscript{22}~W/cm\textsuperscript{2} peak intensity
is shown in Fig.~\ref{fig:eli-pulse}. If there are no cavity reflections,
the foil is rapidly accelerated over a very short distance of 22~$\mu$m,
achieving the same energy as in the case of picosecond pulse, and then ``sails away''
at a constant speed. If reflections at the inner cavity wall are taken
into account (with $R_{c}=0.7$), the foil receives additional kick
by the redirected pulse, which re-interacts with the foil at the distance
of $113$~$\mu$m from its initial position. After being
reflected by the foil for the second time the pulse is redirected
onto the foil once again and hits the moving foil for the third time
at the distance of $645$~$\mu$m from its initial position, but
the effect of this third impact is rather small.  For $R_{c}=0.5$ the energies of the
accelerated ions are slightly smaller, but the general pattern is
the same.

Interestingly, energies achieved over the acceleration distance of
several hundred micrometers in this scenario are very close to energies
achieved with the picosecond pulse, which indicates that reflections of very high order
have small effect on
the resulting ion energies when ultra-thin target foils are used.
This is an important information for the design of cavity targets,
because it shows that it is enough for the cavity reflector to survive
only few reflections, since the loss of further reflections due to
a deteriorating cavity wall would not significantly affect the final
result. However, this picture could be different if a more dense target is
used, for example a metal foil. Such a foil would accelerate less rapidly,
so even for a small acceleration distance many cavity reflections
could occur. The ion energy per a.m.u. achieved without cavity reflections
would be small in this case, but the enhancement due to presence of
the cavity could be much bigger than 100\%.

Figure~\ref{fig:eli-pulse} illustrates one possible risk for cavity-enhanced acceleration in the setup discussed above: the accelerated foil performs "free flight" over extended distances, because the spatial extension of the laser pulse is short compared to cavity depth and there is no overlap between the original pulse and redirected reflections. There is risk that during such motion plasma would expand and its density would drop below critical density, making it effectively transparent to the laser light. This could be avoided by reducing the cavity depth, as much as it is allowed by technological constraints. The effect of changing $L_{c}$ from $80$~$\mu$m to $60$~$\mu$m is illustrated in Fig.~\ref{fig:eli-pulse} with a dotted line.


\section{Conclusions}
Summarizing, we presented a simple implementation of the cavity-enhanced laser-driven ion acceleration - dubbed LICPA - in the light-sail regime.  The model accomodates in a
general way the additional contribution from pulses reflected from
the target foil and redirected towards the target via reflection at
the inner cavity wall. We had shown how the effect of these reflections may be described by a semi-analytic formula. 
We used this formula to obtain concrete predictions
for acceleration of an ultra-thin carbon foil, driven by laser pulses 
of different duration - a ps range pulse and  a fs range pulse - and
different intensity, but the same laser fluence, in the parameter range that would become available at forthcoming facilities, providing a convincing
illustration of characteristic features of LICPA.   The potential 
for a considerable increase in the ion energy and the laser-to-ion
conversion efficiency relative to conventional laser-driven ion acceleration - by a factor of $1.5$ or more - was clearly shown.
Our results provide additional support for the use of circularly polarized beams for laser-driven acceleration. In addition to the fact that the basic interaction of such beams with solid targets leads to smaller target heating and smaller energy spread of accelerated ions, the fact that a substantial part of the beam is reflected from the target makes it possible to enhance the acceleration by recirculating the reflected pulse.

This work was supported in part by the Polish National Science Centre under grant no.\ 871-1/N-SILMI-RNP/2010/0. It was also partially supported by the European Atomic Energy Community under the contract of Association between EURATOM and the Institute of Plasma Physics and Laser Microfusion no.\ FU07-CT-2007-00061 and is subject to the provisions of the European Fusion Development Agreement.


%
%

\end{document}